\documentclass[a4paper]{jpconf}
\usepackage{graphicx, import}
\usepackage[verbose]{placeins}
\usepackage[usenames]{color}
\usepackage{transparent}
\usepackage{caption}

\FloatBarrier
\begin{document}
\title{Passive-performance, analysis, and upgrades of a 1-ton seismic attenuation system}

\author{G Bergmann$^1$, C M Mow-Lowry$^{1,4}$, V B Adya$^2$, A Bertolini$^5$, M M Hanke$^1$, R Kirchhoff$^1$, S M K\"ohlenbeck$^1$, G K\"uhn$^1$, P Oppermann$^1$ A Wanner$^2$, T Westphal$^1$, J W\"ohler$^1$, D S Wu$^1$, H L\"uck$^{2}$, K A Strain$^3$, K Danzmann$^{1,2 }$}

\address{$^1$ Albert-Einstein-Institut, Max-Planck-Institut f\"ur Gravitationsphysik, D-30167 Hannover, Germany}
\address{$^2$ Leibniz Universit\"at Hannover, D-30167 Hannover, Germany}
\address{$^3$ SUPA, School of Physics and Astronomy, University of Glasgow, G12 8QQ, UK}
\address{$^4$ University of Birmingham, Birmingham B15 2TT, United Kingdom}
\address{$^5$ Nikhef, Science Park, 1098 XG Amsterdam, Netherlands}

\ead{gerald.bergmann@aei.mpg.de}

\begin{abstract}
The \textit{10\,m Prototype} facility at the Albert-Einstein-Institute (AEI) in Hanover, Germany, employs three large seismic attenuation systems to reduce mechanical motion. The AEI Seismic-Attenuation-System (AEI-SAS) uses mechanical anti-springs in order to achieve resonance frequencies below 0.5\,Hz. This system provides passive isolation from ground motion by a factor of about 400 in the horizontal direction at 4\,Hz and in the vertical direction at 9\,Hz. The presented isolation performance is measured under vacuum conditions using a combination of commercial and custom-made inertial sensors. Detailed analysis of this performance led to the design and implementation of tuned dampers to mitigate the effect of the unavoidable higher order modes of the system. These dampers reduce RMS motion substantially in the frequency range between 10 and 100\,Hz in 6 degrees of freedom. The results presented here demonstrate that the AEI-SAS provides substantial passive isolation at all the fundamental mirror-suspension resonances.
\end{abstract}

\section{Introduction}
The AEI 10\,m Prototype was designed to provide a low-noise environment for testing gravitational-wave detector technology and to probe the limits of high-precision interferometry~\cite{gossler2010aei}. Currently, an interferometer to reach and surpass the Standard Quantum Limit (SQL) with 100\,g mirror masses is being installed~\cite{buonanno2001optical}. In parallel, novel techniques are being developed and tested both for upgrades to advanced gravitational-wave detectors (Advanced LIGO~\cite{aasi2015advanced}, Advanced Virgo~\cite{TheVirgo:2014hva}, and KAGRA~\cite{aso2013interferometer}) as well as for future observatories such as the Einstein Telescope~\cite{punturo2010einstein}.
\\A crucial element of the 10\,m Prototype design is isolation from environmental vibrations. A large vacuum envelope (100\,m$^3$, $10^{-7}\,$hPa) houses the entire apparatus and shields against acoustic coupling. Seismic Attenuation Systems (SASs) isolate all essential components from vibrations caused by seismic and anthropogenic ground motion, and from structural acoustics. Two SASs are installed and operating and a third SAS is currently being used to test mechanical design upgrades.

The $1.75\,{\rm m} \times 1.75\,{\rm m}$ optical table isolated by the AEI-SAS offers a spacious platform to perform various high precision experiments. In particular, the design of the mirror suspensions for the SQL-interferometer and a reference cavity \cite{futhesis} rely on pre-isolation at their fundamental resonances (between 0.6\,Hz and 27\,Hz). This isolation of the suspension point reduces the total actuation forces required to control the suspended mirrors, enabling the use of low-noise actuators that meet the SQL-interferometer's noise requirements. 

In this paper, we present the passive performance of the AEI-SAS, compare it with a simple analytical model, and discuss the observed differences. These are mainly due to cross-coupling, at low frequencies, and internal resonances, at high frequencies. This knowledge was used to improve the isolation performance by damping and shifting the internal resonances. 

An adapted version of the AEI-SAS, the External Injection Bench Seismic Attenuation System (EIB-SAS), is used to isolate injection optics at Advanced Virgo \cite{blom2015vertical}. The EIB-SAS has higher fundamental resonant frequencies compared with the AEI-SAS, allowable due to the relatively relaxed noise requirement. This increases stability at the expense of isolation performance. 

Advanced LIGO's in-vacuum seismic isolation relies on an active isolation feedback to reduce motion below about 5\,Hz. The single-stage Internal Seismic Isolators (HAM-ISI) is most comparable with the AEI-SAS.

Despite the difference in design philosophies and performance requirements, it is possible to compare figures~\ref{vperformance} and \ref{hperformance} of this paper with figures 16 and 17 in reference~\cite{matichard2015seismic}. 
 Active feedback will significantly improve the AEI-SAS performance around its fundamental resonances and will be subject of a future publication.

\section{Mechanical design of the AEI-SAS}
The AEI-SAS is composed of two stages: a horizontal isolation stage based on three Inverted-Pendulum (IP) legs, and a vertical isolation stage with three Geometric Anti-Spring (GAS) filters~\cite{wanner2012seismic}.

\begin{figure}[!htb]
	\centering
	\includegraphics[width=0.85\textwidth]{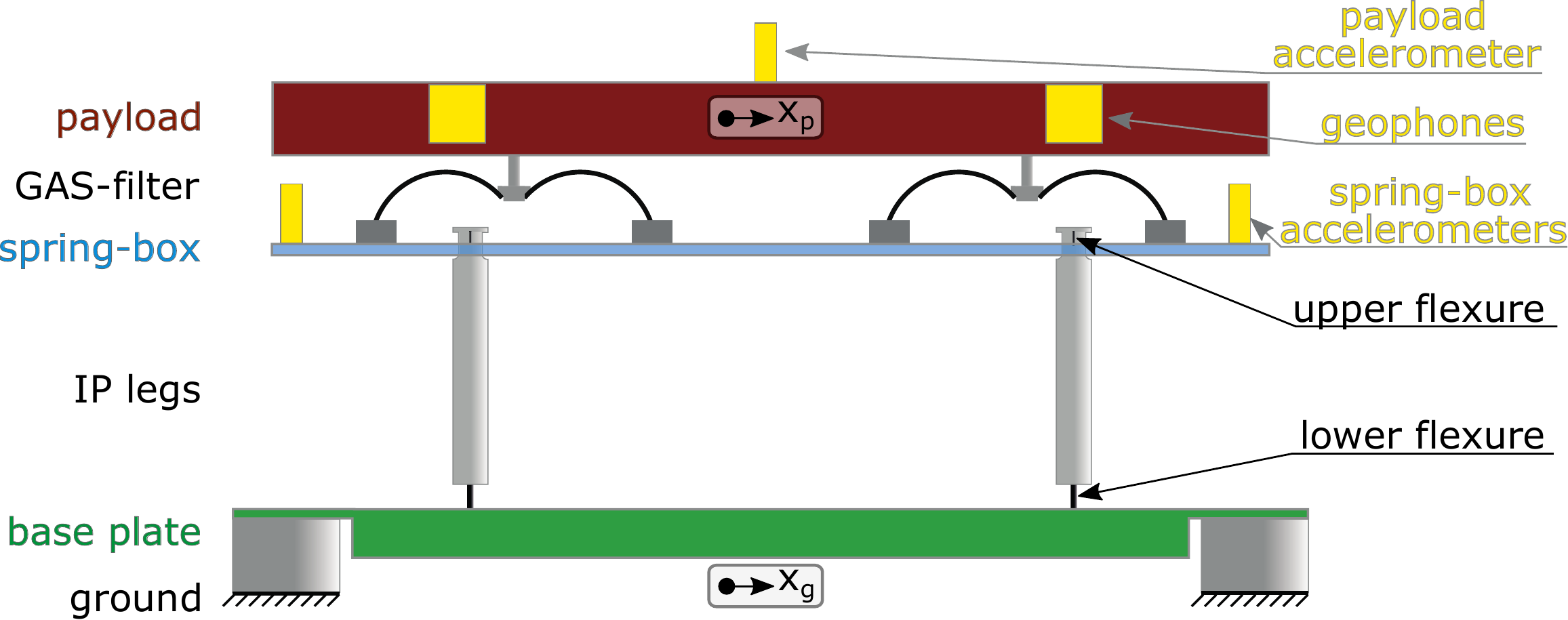}
	\caption{Simplified sketch of the original  AEI-SAS design: The payload is supported by three GAS-filters providing vertical isolation. The diagram shows an unfolded version, but in reality the GAS filters are nested inside an aluminum structure, the spring-box, within the horizontal stage (see figure \ref{CADsection}). The spring-box is supported by three IP-legs providing horizontal isolation. The bottom of the IP-legs are connected via flexures to the baseplate. This baseplate is rigidly connected to the `feet' of the vacuum tank. Three vertical geophones are installed inside the payload Horizontal motion is sensed by three custom made accelerometers~\cite{bertolini2006mechanical} that are placed on the springbox. An auxiliary horizontal accelerometer is installed on top of the payload. }
	\label{tablediagram}
\end{figure}

\begin{figure}[!htb]
	\centering
	\includegraphics[width=0.85\textwidth]{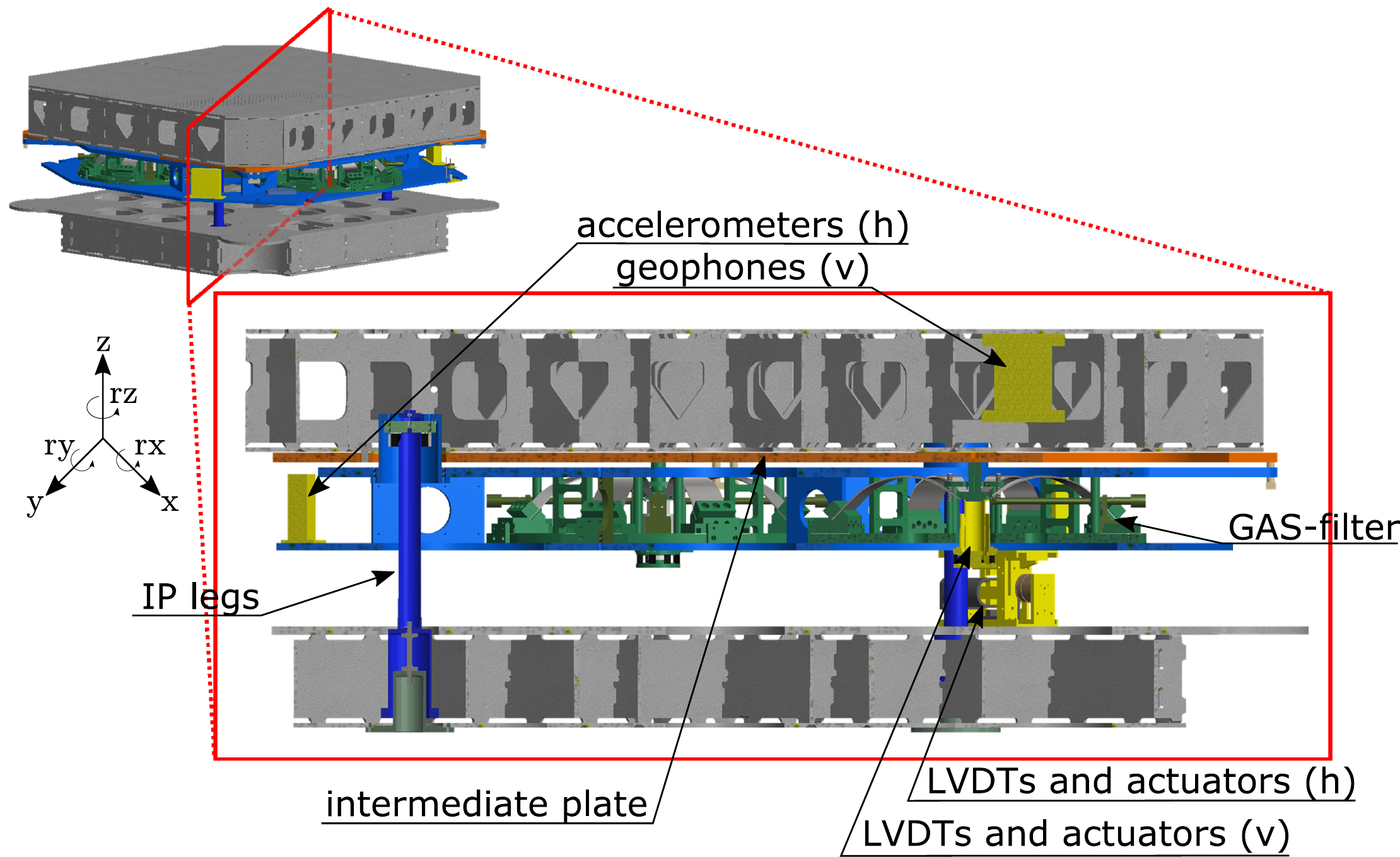}
	\caption{CAD model of the AEI-SAS: A section view shows the structure of the individual isolation stages and the location of the single components. For simplicity componants as the motorized springs and the tilt stabilization are not shown in this model. This figure shows the original version of the AEI-SAS with the old IP-legs and a rigid intermediate plate- payload connection. }
	\label{CADsection}
\end{figure}

Each SAS supports a payload of approximately 900\,kg. This includes a 530\,kg, $1.75\,{\rm m} \times 1.75\,$m stainless steel breadboard with an internal honeycomb structure. The breadboard is mounted on the three GAS filters. Each GAS filter consists of a crown-shaped assembly of eight maraging steel blades \cite{Stochino2007}. These blades are bent and fastened to a central key-stone that supports the table top. Compression of opposing blades against each other introduces negative stiffness along the vertical axis, lowering the effective spring constant~\cite{bertolini1999seismic, cella2005monolithic}. 

The spring-box is suspended by means of $3\,$mm thick and $25\,$mm long cylindrical maraging-steel flexures connected to the top of the IP legs on their other side. The leg itself is an aluminum tube with $1\,$mm wall thickness. The horizontal restoring force is provided by a $10.6\,$mm thick and $60\,$mm long maraging-steel flexure connecting the IP leg to the baseplate at the bottom (see figure \ref{tablediagram} and \ref{CADsection}). The SAS baseplate is bolted to the inside of the rigid `feet' of the vacuum tank. On the outside, these feet are bolted to the 80\,cm thick concrete foundation of the lab. 
The AEI-SAS provides passive vibration isolation for the payload above the fundamental resonance frequencies. Additionally, several sensors and actuators are installed to measure and control the payload position actively. Three commercial L-22D geophones~\cite{sercel}, arranged in an equilateral triangle, measure the inertial vertical payload motion. Three custom made accelerometers~\cite{bertolini2006mechanical} sense the inertial horizontal motion of the system. They are placed close to the edges of the spring-box and are also arranged in an equilateral triangle.

The displacement between the ground and the payload is measured by Linear Variable Differential Transformers (LVDTs)~\cite{tariq2002linear}. There are three vertical and three horizontal LVDTs. Each one is co-located with a voice-coil actuator. All table control forces are applied using these actuators. A commercial STS-2 triaxial broadband seismometer is used to measure the ground motion of the laboratory floor. 

Above the individual resonance frequencies, the AEI-SAS decouples the payload from ground motion in 6 degrees of freedom. Table 1 shows the resonance frequencies of the AEI-SAS and the corresponding masses \cite{wanner2013seismic} and moments of inertia. The displayed values vary during commissioning of the system. The resonance frequencies of the horizontal degrees of freedom are tuned by mass adjustment.  All three moments of inertia depend on the mass distribution of the payload, which will vary, due to changing experiments in the 10\,m Prototype. Fairly high tilt stiffness is ensured in rx and ry for stability reasons. 

\begin{table}[]
\centering
\label{tabel_res}
\caption{Resonance frequencies, and corresponding masses \cite{wanner2013seismic} and moments of inertia of the AEI-SAS. Figure~\ref{CADsection} shows the corresponding coordinate system. The moments of inertia are determined with respect to the principal axes of either the payload (for rx and ry) or the payload and the spring-box (rz). Note that the rx and ry axes do not coincide with the principle axes of the SAS.   }   
\renewcommand{\arraystretch}{1.3}

\begin{tabular}{ccc}
Direction & Resonance frequency  {[}Hz{]} & Mass {[}kg{]}                                   \\  \hline
x         & 0.16                          & 1231                                            \\ 
y         & 0.16                          & 1231                                            \\ 
z         & 0.27                          & 900                                             \\ 
          &                               & Moment of inertia  {[}kg m$^2${]}								 \\  \hline
rx        & 0.4                           & 260                                             \\ 
ry        & 0.4                           & 260                                             \\ 
rz        & 0.1                           & 598                                             \\ \hline
\end{tabular}
\end{table}

A more detailed description of the AEI-SAS and its instruments can be found in Wanner {\it et al.}~\cite{wanner2012seismic}.

\section{Passive isolation performance}
\begin{figure}[!htb]
	\centering
	\includegraphics[width=0.85\textwidth]{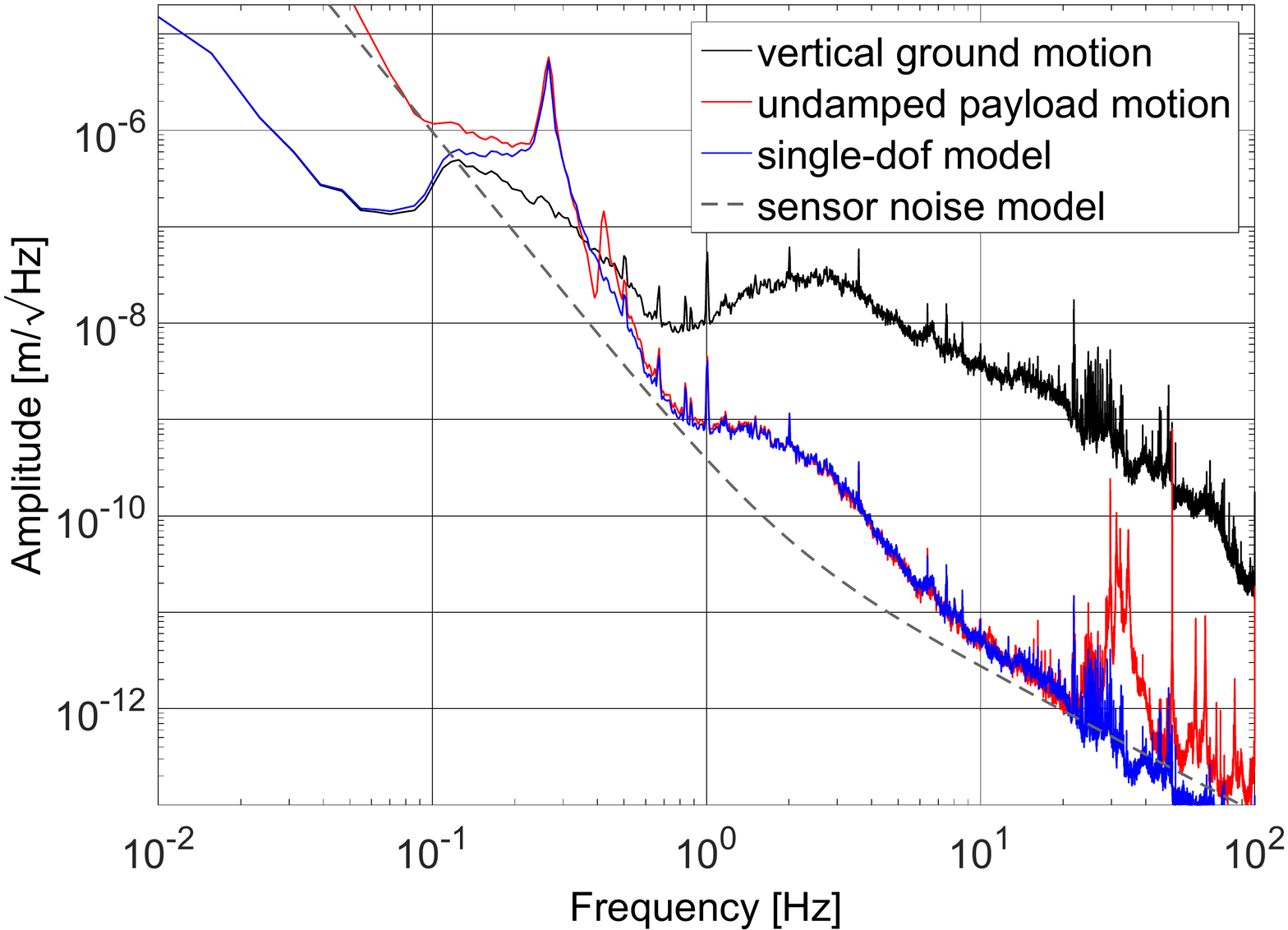}
	\caption{Vertical passive AEI-SAS performance: A comparison of the measured table-top motion (red) and the predicted table-top motion (blue) based on the ground motion (black). The model is a simple harmonic oscillator with a fitted resonance frequency, quality factor, and CoP plateau. Above $30\,$Hz the sensor noise (gray) is dominant in the measurement. The relative deviation of measurement and predicted table motion averaged across all measurement points between $0.1\,$Hz and $20\,$Hz is less than $4\,$\%. The internal resonances with the lowest frequencies (described in section~\ref{sectionspringboxres}) are the spring-box resonances between $30$ and $40\,$Hz.}
	\label{vperformance}
\end{figure}

\begin{figure}[!htb]
	\centering
	\includegraphics[width=0.85\textwidth]{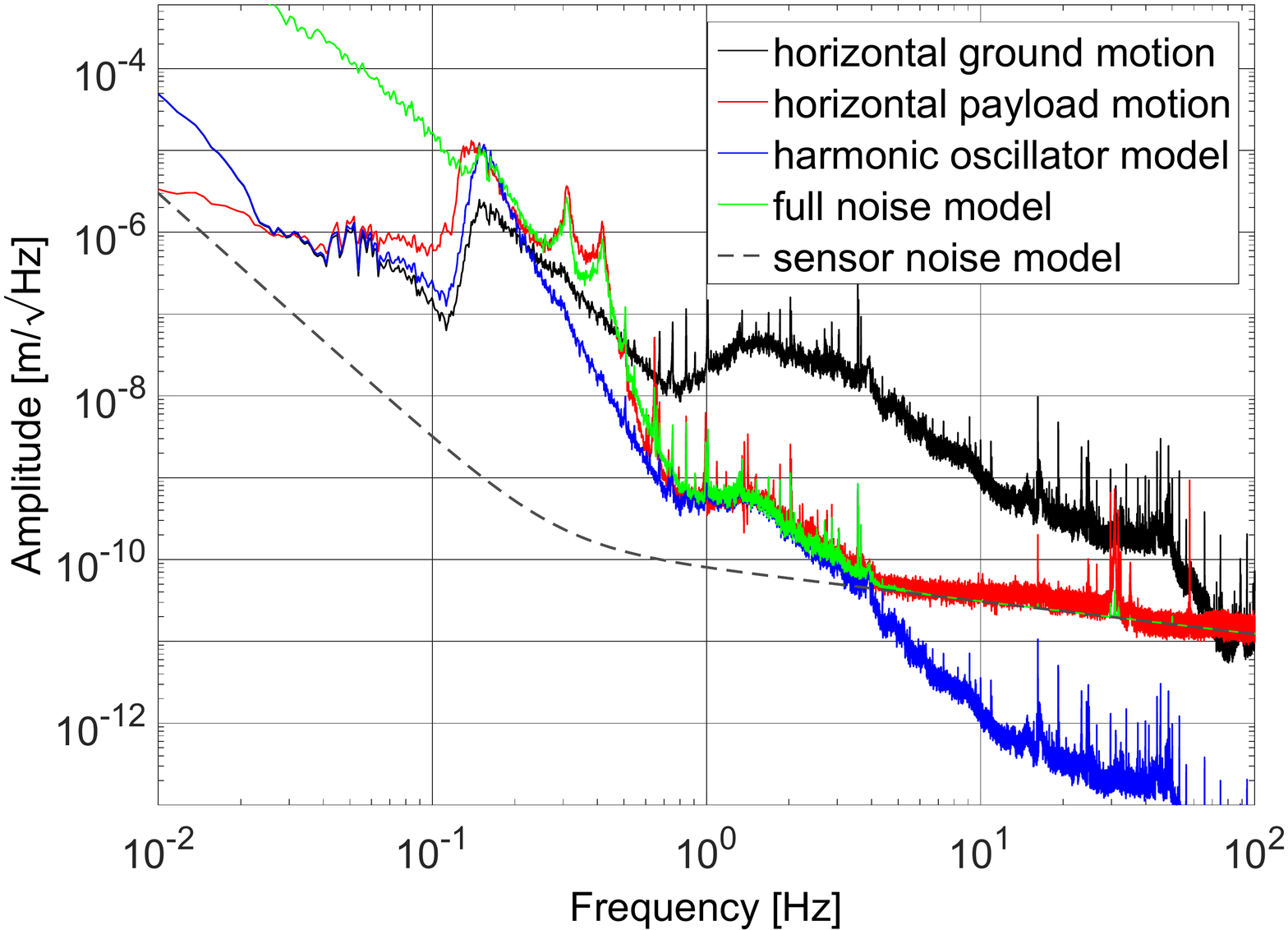}
	\caption{Horizontal passive AEI-SAS performance: a comparison of the measured table-top motion (red), the single-dof model of the table-top motion (blue), and the ground motion (black). The model is again a simple, massive harmonic oscillator model multiplied by the ground motion. A full noise model (green) includes cross-coupling from the payload tilt, dominant at low frequencies, and the accelerometer sensor noise, dominant above 5\,Hz. A model of the accelerometer noise is shown in gray. The average difference between the measured and predicted motion between $0.1\,$Hz and $5\,$Hz is less than $1\,$\%.}
	\label{hperformance}
\end{figure}

The passive performance of the AEI-SAS is shown in figure \ref{vperformance} and \ref{hperformance} in red. This measurement is compared to a simple one-dimensional model, which is explained in the following paragraph.

A good measure of isolation performance is the transmissibility, $x_\mathrm p/x_\mathrm g$, of ground motion, $x_\mathrm g$, to payload motion, $x_\mathrm p$. It can be deduced from the equation of motion of the system. Both the horizontal and vertical stages can be modeled as a harmonic oscillator with finite mass with 
\begin{equation}
	\omega_\mathrm n ^2  \left(x_\mathrm p-x_\mathrm g \right) + \beta \  \ddot{x}_\mathrm g + \ddot{x}_\mathrm p =0 \enspace,
	\label{equofmotion}
\end{equation}
where  $\omega_\mathrm n$ is the complex natural frequency, related to the fundamental resonance frequency $\omega_\mathrm 0$ and the loss angle of the spring material $\phi_\mathrm h$ by $\omega_\mathrm n^2=\omega_ 0^2(1+ i \phi_\mathrm h)$. The bypass $\beta$ is a function of the IP-legs' (or GAS filter blades') geometry and mass distribution, and the payload mass. It determines the height of the Center of Percussion (CoP) plateau~\cite{losurdo1999inverted, stochino2007improvement}. The mechanics of the CoP tuning are described in ~\cite{wanner2013seismic} in more detail.  
The transmissibility from ground motion to payload position follows from the Fourier transform of equation~\ref{equofmotion}
\begin{equation}
	T(\omega) = \frac{x_\mathrm p(\omega)}{x_\mathrm g(\omega)} =\frac{\omega_\mathrm n ^2+\beta \omega^2}{\omega_\mathrm n ^2-\omega^2} \enspace.
\end{equation}

To generate a vertical transmissibility function, the resonance frequency, $f_\mathrm{0v}=0.27\,$Hz, damping factor $\phi_\mathrm v=\frac{1}{30}$, and CoP factor, $\beta_\mathrm v=7\times 10^{-4}$, were fitted to the measured vertical payload motion (the red curve in figure~\ref{vperformance}). The vertical ground motion (the black curve) was then multiplied by the vertical transmissibility function $T_\mathrm v$ to produce the blue curve, a single-dof model of the vertical payload motion.  The sensor noise of the three vertical L-22D ~\cite{sercel} installed in the payload dominates the measurement above $\sim 30\,$Hz. The figure shows the passive performance, free from position control or feedback forces, and it was recorded with the vacuum system pumped down to a pressure below $10^{-5}\,$hPa. Up to the first internal resonances above 30\,Hz, the measurement matches the predicted payload motion very well. The small peak at 0.4\,Hz is due to cross-coupling from the fundamental tilt resonance.

The horizontal payload motion in figure~\ref{hperformance} (red) was measured by an auxiliary horizontal accelerometer placed on the top of the payload. It is compared with the ideal theoretical horizontal table motion (blue). The transmissibility of a harmonic oscillator $T_\mathrm h$  having a fundamental resonance at $f_\mathrm{0h}=0.16\,$Hz, a damping factor of $\phi_\mathrm h=\frac{1}{5}$, and a center of percussion factor of $\beta_\mathrm h=10^{-4}$ is multiplied by the ground motion (black). At low frequencies the measurement differs from the model due to coupling between payload tilt and the accelerometer readout. The tilt of the payload by an angle $\Theta$ is seen by the accelerometer as a translation of  $x_\mathrm t=\Theta \times g/\omega^2$. That is because of tilt-horizontal coupling, where gravity is assumed constant, and cannot separate tilt and acceleration with a single instrument~\cite{matichard2015review}. 
\begin{figure}[!htb]
	\centering
	\includegraphics[width=0.85\textwidth]{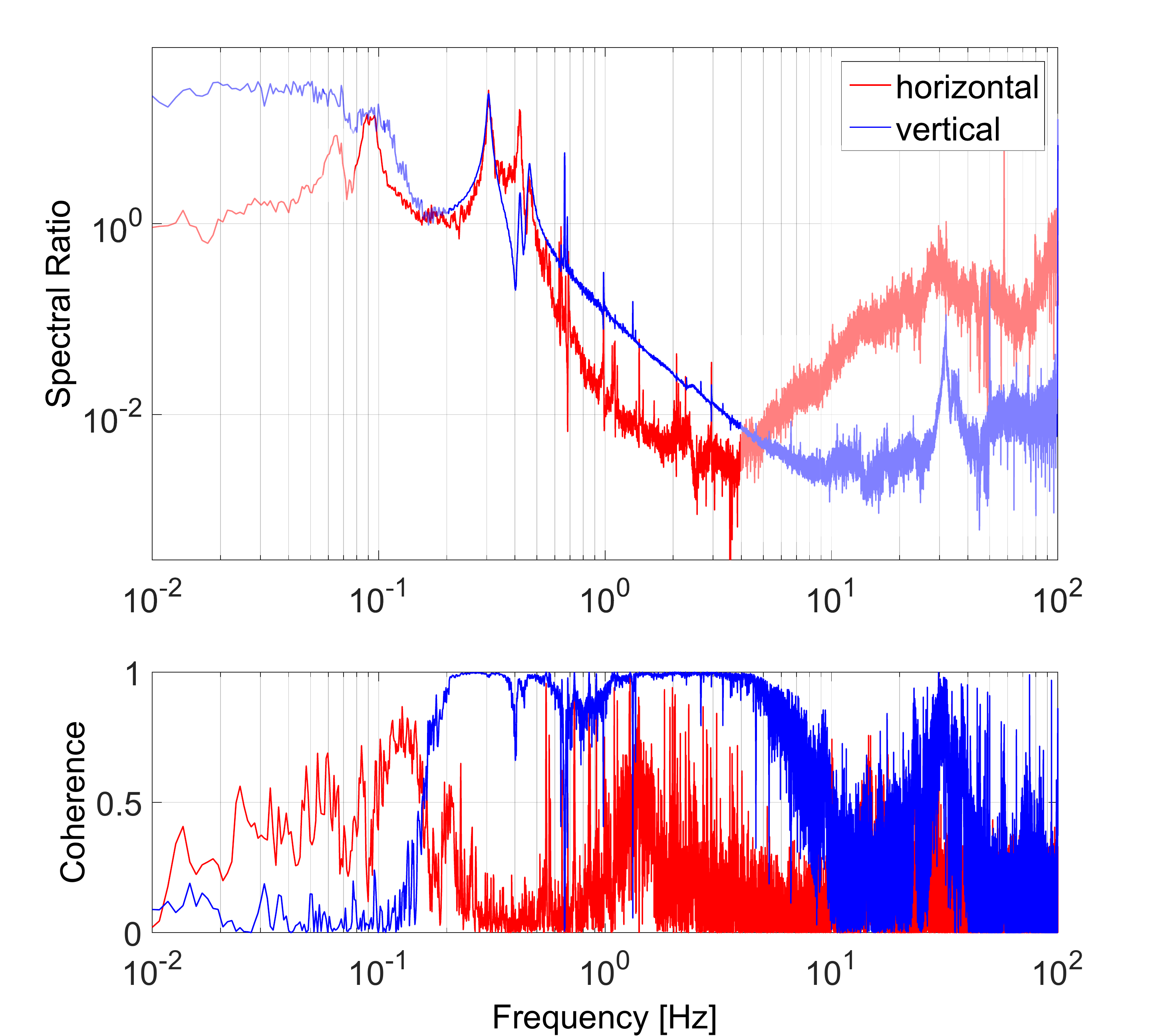}
	\caption{Spectral ratio and coherence of ground motion to payload motion: Compared to figure~\ref{vperformance}, the ground motion at the time of this measurement was lower, so sensor noise is limiting the vertical measurement (blue) below $\sim0.2\,$Hz and above $\sim 9\,$Hz. The horizontal measurement (red) is limited by sensor noise below $\sim 0.08\,$Hz and above $\sim 4\,$Hz. The sharp peaks around 1 Hz are due to the recoil of the fundamental modes of the mirror suspensions, which were already installed on top of the AEI-SAS at the time of this measurement. At 4\,Hz the horizontal payload motion is $2.6 \times 10^{-3}$ lower than the ground motion. The vertical payload motion can be measured up to 9\,Hz, where it is $2.4 \times 10^{-3}$ times lower than the ground motion in this direction. The data in this figure was recorded after the Fluorel pads and the spring-box damper (described in section~\ref{sectionfluorel} and~\ref{springboxdamper}) were installed. }
	\label{inchtr}
	\end{figure}
This tilt-horizontal coupling at low frequencies as well as the sensor noise at high frequencies is included in the full noise model N (green) in figure~\ref{hperformance}.
\begin{equation}
N^2=\left(T_\mathrm h x_\mathrm g \right)^2+\left( \Theta_\mathrm p \frac{g}{\left(2 \pi f \right)^2} \right)^2+n_\mathrm s ^2
\end{equation}
 The payload tilt motion $\Theta_\mathrm p$ was measured using differential vertical signals from the L-22D geophones mounted in the payload. Low-frequency noise in the geophones causes the strong deviation of the green trace below $0.15\,$Hz. Above 5\,Hz the green curve follows the measured sensor noise $n_\mathrm s$ of the accelerometers. As with the vertical payload motion, internal resonances show up above 30\,Hz. These resonances are discussed in more detail in the following section. Note that the measurements in figures~\ref{vperformance}, figure~\ref{hperformance}, and figure~\ref{inchtr} were made after the installation of Fluorel pads described in section~\ref{sectionfluorel}, and as such the 17\,Hz resonance is not present in this data.

Additional information about the overall passive performance of the AEI-SAS can be obtained by comparing  payload motion to ground motion. Figure~\ref{inchtr} shows the spectral ratio $\left|x_\mathrm p\right|/\left|x_\mathrm g\right|$. The coherence of $x_\mathrm p$ and $x_\mathrm g$ in the lower graph shows in which frequency band the sensor signal is caused by the ground motion in the same direction. The ground motion sensor is not limited by sensor noise at any frequency where the payload sensors provide good signals. In the horizontal direction, the low coherence between 0.3 \,Hz and 1\,Hz is caused by cross coupling from other degrees of freedom. Below $0.08\,$Hz and above $4\,$Hz the payload measurement is limited by sensor noise. Similar to the measurement in figure~\ref{vperformance} and figure~\ref{hperformance}, low-frequency payload tilt couples strongly in to the horizontal sensors. In the vertical direction sensor noise limits the measurement below $0.2\,$Hz and above $9\,$Hz. 

The data in figure~\ref{inchtr} shows that the payload motion is $2.6 \times 10^{-3}$ times lower than the ground motion in the horizontal direction at 4\,Hz, and  $2.4 \times 10^{-3}$ times lower than ground motion in the vertical direction at 9\,Hz. Above those frequencies the measurement is limited by measurement noise, but based on prior  measurements of the driven transmissibility, we expect more attenuation at higher frequencies.  In a shaker test-stand, where the baseplate was excited in the horizontal direction, a peak horizontal isolation of $10^{-4}$ at 7\,Hz was achieved. A maximal vertical isolation of $10^{-4}$ above 20\,Hz was measured using a single GAS-filter~\cite{wanner2012seismic}.

The lowest horizontal resonance frequency of the mirror suspensions installed on the AEI-SAS is 0.64\,Hz. At this frequency the horizontal payload motion is already 12 times lower than ground motion. At the lowest vertical mirror suspension resonance of 1\,Hz, the vertical payload motion is 8 times lower than ground motion.

\section{Analysis and mitigation of horizontal modes}
As shown in figure \ref{vperformance} the AEI-SAS follows the physics of a massive harmonic oscillator up to a frequency where the individual components are not ideally stiff, but instead behave as additional spring-mass systems. The origin of these internal resonances and possible improvements to the AEI-SAS are discussed. In the following, our efforts are focused on parasitic resonances below 60\,Hz, where ground motion suppression is required in order to reduce the excitation of the fundamental resonances of the payload, in particular the mirror suspensions.

The model used in the previous section, based on the assumption that the spring-box is massless and that horizontal and vertical suspension stages are totally decoupled, is too simple. A more realistic description of the system must take into account that:
\begin{itemize}
\item the CoM of the payload is located well above the CoM of the spring-box. This determines the observed large coupling between horizontal and tilt degrees of freedom,
\item the IP legs have a finite vertical compliance, and
\item the GAS filters have a finite horizontal compliance.
\end{itemize}
In particular, the parasitic compliance of the suspension elements, combined with the sizeable mass (about 300\,kg) of the spring-box, is such that the 6 rigid-body modes of the spring-box lie in the 10\,-\,50\,Hz frequency band. This affects the overall attenuation performance of the system significantly. The large mass ratio between spring-box and payload (about 1\,:\,3) causes a large transmissibility of ground vibrations around the spring-box parasitic modes. In the following section, these modes are characterized and efforts to mitigate their impact are presented.

\subsection{Testing and modeling of horizontal modes}
The parasitic modes of the isolation system with the lowest frequency are differential modes between the spring-box and the payload, both in horizontal translation and rotation. These modes show up at about 17\,Hz and were investigated experimentally and with simulations.

\begin{figure}[!htb]
	\centering
	\includegraphics[width=0.85\textwidth]{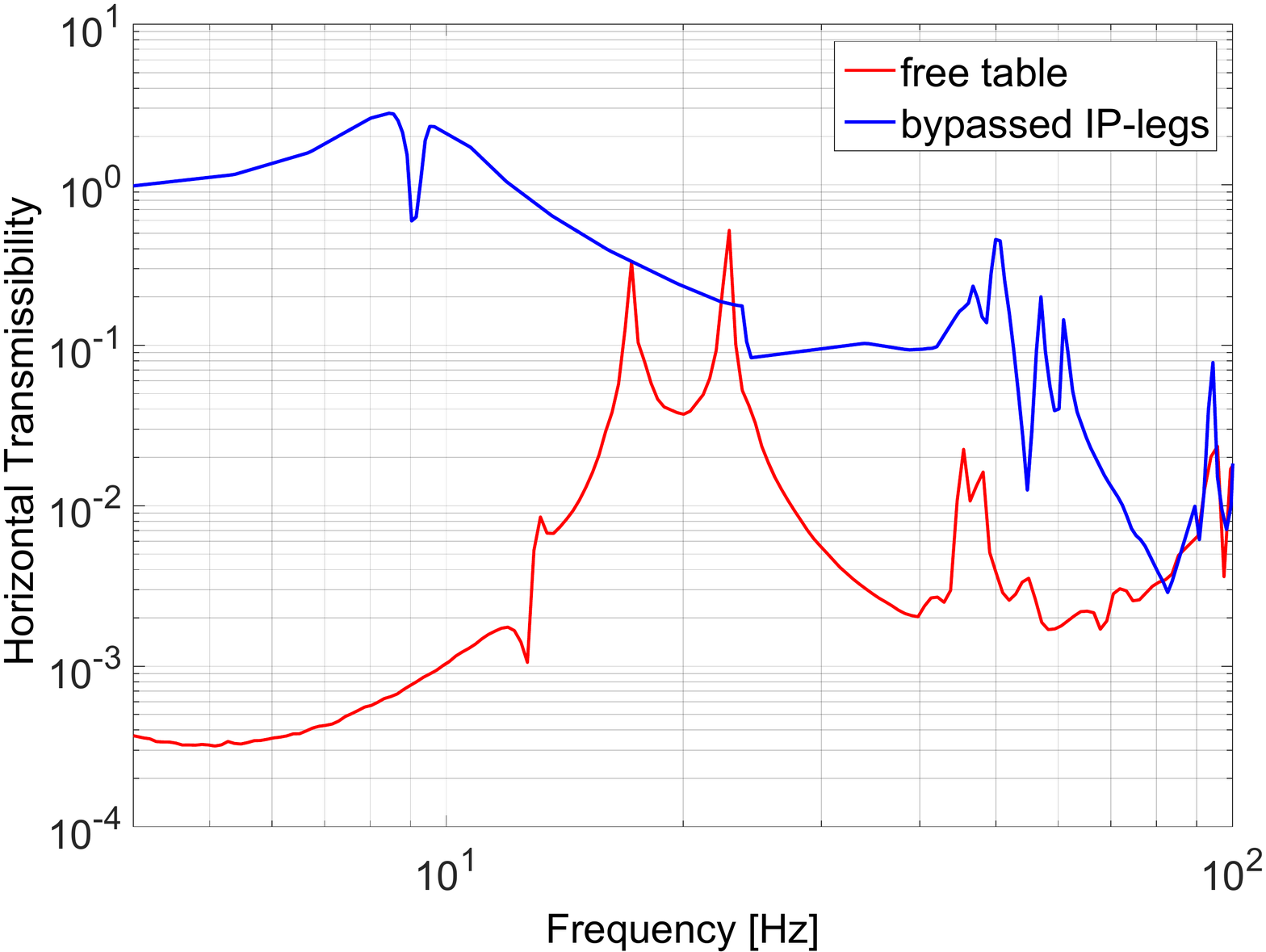}
	\caption{Investigation of internal resonances using the test-stand: The SAS's baseplate is driven horizontally, and the transmissibility from the base plate to the payload is shown for normal operation (red) and with the horizontal stage clamped (blue). The 9\,Hz structure in the blue curve is the spring-box mode caused by the horizontal compliance of the GAS-filters. In the free table (red curve), the horizontal compliance of the GAS filters causes differential oscillations between spring-box and payload at 13\,Hz and 17\,Hz. Finite element simulations suggest, that the 22\,Hz resonance peak is due to a bending mode of the intermediate plate.
	}
	\label{shaker}
\end{figure}

In order to measure and tune the SAS performance, the whole 1 ton table was installed in a shaker test-stand. The SAS's baseplate was suspended from four wires so that the SAS could freely move in the horizontal degrees of freedom. A horizontal voice coil actuator was used to apply translational force to the baseplate, enabling fast and coherent measurements of the isolation system's transmissibility. Figure~\ref{shaker} shows the horizontal transmissibility from the base plate to the payload (red). For comparison the transmissibility was re-measured when the spring-box was clamped to the baseplate, effectively bypassing the IP-legs (blue).

When the spring-box is clamped, the softest connection in the horizontal direction is the horizontal compliance of the three GAS-filters. The free SAS has its lowest horizontal internal resonance at 17\,Hz. For the clamped system it shifts to about $9\,$Hz. Considering this setup as a simplified one-dimensional harmonic oscillator, where the 932\,kg payload is connected by a spring to the ground, its effective spring constant is $\kappa_{hGAS} (9\,$Hz$)=3\times10^6\,$N/m. The free AEI-SAS would then correspond to a simplified one-dimensional model, where the payload mass is connected via the same spring to the spring-box mass of 331\,kg. Neglecting the horizontal IP-leg stiffness, this system has its resonance frequency at 17.5\,Hz. 

\begin{figure}[!htb]
	\centering
	\includegraphics[width=0.75\textwidth]{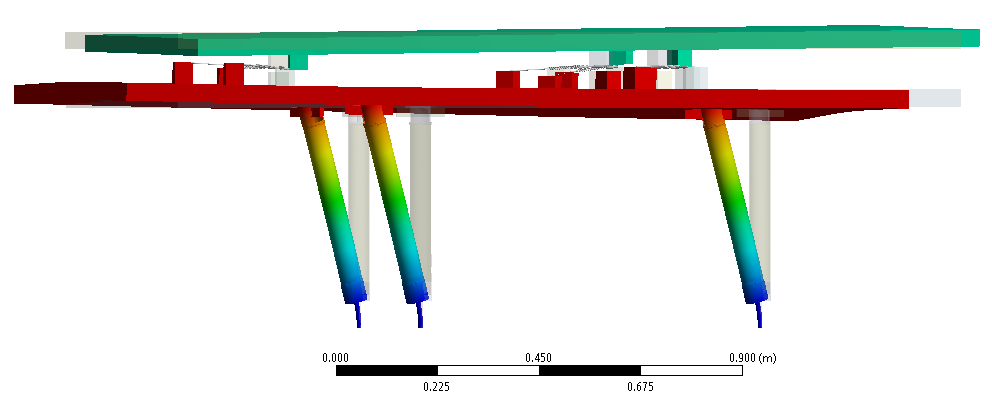}
	\caption{Finite element simulation of the horizontal GAS-filter compliance: The spring-box and the payload are represented by stiff plates. The GAS-filters are modelled as a set of springs with the measured horizontal and vertical stiffness. The resulting modes are translational and rotational differential oscillations between spring-box and payload.}
	\label{fem}
\end{figure}

An Ansys Workbench finite element simulation (figure~\ref{fem}) shows the resulting mode shapes. The  model is a simplified version of the AEI-SAS with the correct horizontal dimensions and masses for both the spring-box and the payload. The model for the IP-legs is close to reality. Vertical dimensions and mass distribution are simplified to produce an efficient mesh. The GAS-filters are modelled as a set of two horizontal springs, using the spring constant calculated above, and one vertical spring with a stiffness of $400\,$N/m, calculated based on the fundamental vertical resonance frequency seen in figure \ref{vperformance}.
The 17\,Hz resonances are horizontal differential oscillations between the spring-box and the payload. The corresponding mode to the differential rotation around the vertical axis is at 12.8\,Hz. This mode is seen at 13\,Hz in the horizontal frequency response in figure~\ref{shaker}.

\subsection{Fluorel stage}
\label{sectionfluorel}
The 17\,Hz horizontal GAS-filter mode is highly undesirable in the $10\,$m prototype interferometer since some of the mirror suspensions have a vertical (bounce) modes very close to this frequency. Any overlap of these resonances would result in strongly enhanced mirror motion. Initially, we attempted to stiffen the GAS-filters in the horizontal direction by installing additional structures. However, this approach was discarded because either the low vertical stiffness of the GAS-filters was compromised or because the functional range of the stiffening structure was too small. A second approach, the inclusion of an additional, well-damped spring-mass stage in the SAS proved to be very successful. Three vacuum compatible rubber pads (Fluorel) were placed between the payload and the `intermediate plate', a $113\,$kg aluminum plate mounted on top of the GAS filters, where previously this plate was rigidly connected to the payload.

\begin{figure}[!htb]
	\centering
	\includegraphics[width=0.80\textwidth]{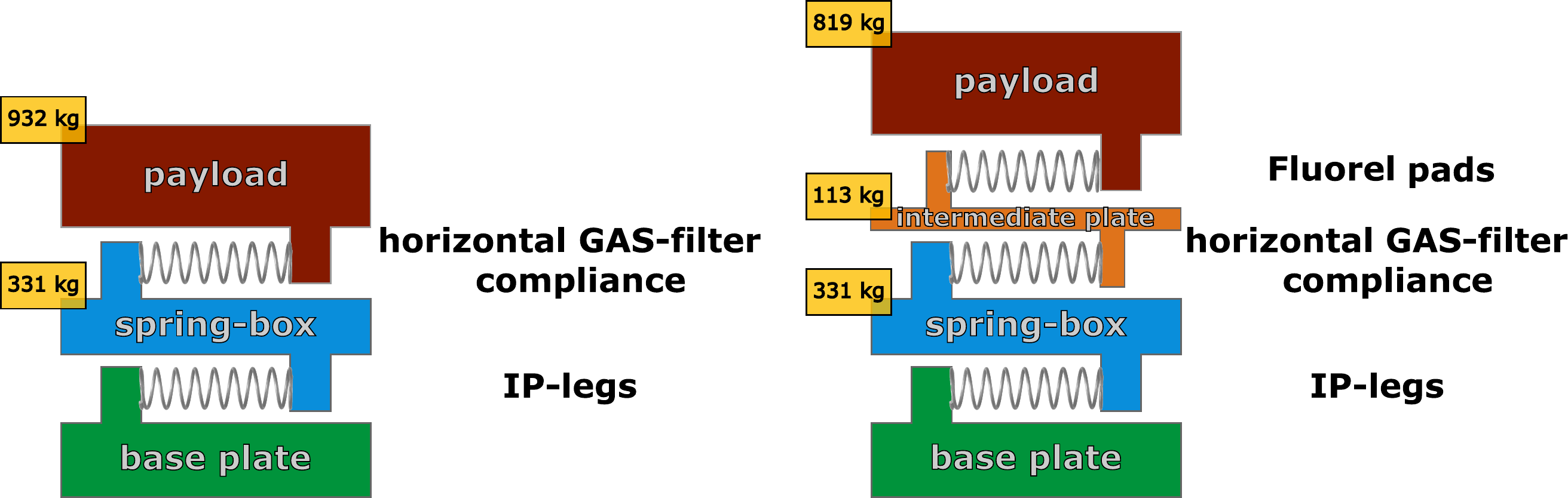}
	\caption{Simplified one-dimensional models of the isolation system show how the implementation of a Fluorel stage changes the horizontal eigenfrequencies.\\
		Left: Original system. The IP-leg spring ($\kappa_\mathrm{ip}\approx 500\,$N/m) connects the spring-box (331\,kg) to the ground. The horizontal GAS-filter spring $\kappa_\mathrm{hGAS}=3\times10^6\,$N/m connects the spring-box to the 932\,kg payload (including the intermediate plate). The common mode of this system is at 0.1\,Hz, the differential mode is at 17\,Hz.\\
		Right: Fluorel pads are placed between intermediate plate and payload. The horizontal stiffness of this spring is about $1.2\times10^6\,$N/m, resulting in eigenfrequencies of 0.1\,Hz, 9\,Hz, and 34\,Hz. Even though one of the internal resonances is at a lower frequency than before, the lossy Fluorel pads strongly damp these oscillations.}
	\label{fluorelp}
\end{figure}

\begin{figure}[!htb]
	\centering
	\includegraphics[width=0.85\textwidth]{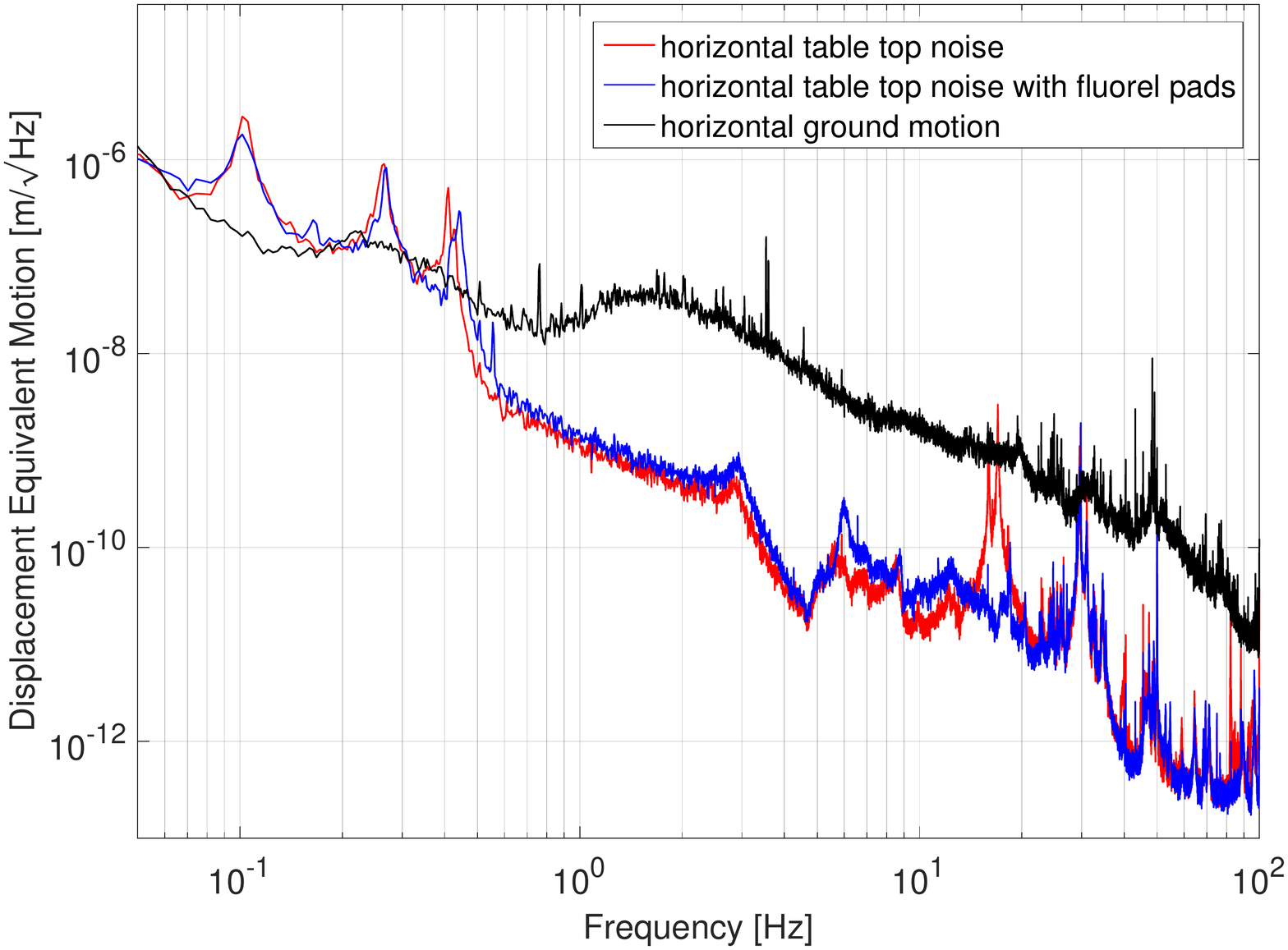}
	\caption{A comparison between horizontal payload motion with (blue) and without (red) the additional Fluorel stage. The strong 17\,Hz resonance is no longer visible. Unlike the spectrum in figure~\ref{hperformance}, this measurement was recorded in air, resulting in excess motion between 1\,Hz and 10\,Hz. The structures from 3-6Hz are not present when measurements are performed under vacuum. The peaks above 30\,Hz are vertical spring-box modes. They are discussed in the following sections. }
	\label{fluorelPSD}
\end{figure}

Figure~\ref{fluorelp} compares the two configurations using simple one-dimensional models. The model for the original design consists of two horizontal springs. One represents the IP-legs with a stiffness of about $\kappa_\mathrm{ip}= 500\,$N/m and the the other the horizontal GAS filter stiffness $\kappa_\mathrm{hGAS}=3\times10^6\,$N/m. The eigenfrequencies of the coupled system are $0.1\,$Hz and 17\,Hz. Implementing the Fluorel stage results in a three-spring, three-mass system. The fundamental mode at 0.1\,Hz is the common mode of the whole system on the IP-leg flexures. At $9\,$Hz the spring-box and the intermediate plate oscillate together between the heavy payload and the ground. A third resonance at 34\,Hz corresponds to the oscillation of the intermediate plate between the payload and the spring-box. Since the lossy Fluorel pads' springs are involved in the latter two modes, they are well damped and do not significantly influence the SAS performance at low frequencies. Figure~\ref{fluorelPSD} shows the substantial performance improvement provided by the additional stage.

\FloatBarrier		
\section{Vertical modes damping}
\label{sectionspringboxres}
\begin{figure}[!htb]
	\centering
	\includegraphics[width=1\textwidth]{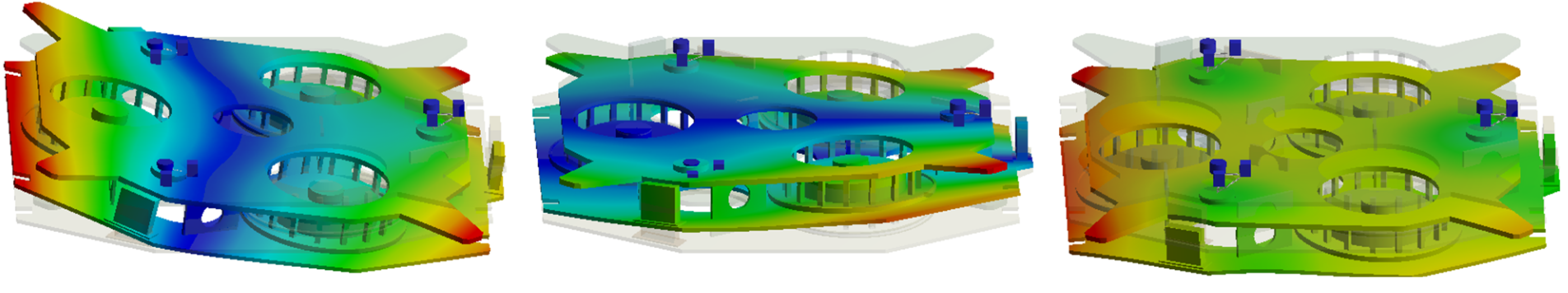}
	\caption{Finite element simulation of the spring-box and IP-leg modes. A set of three springs represent each IP-leg's stiffness matrix. The mass and dimensions of the spring-box are close to reality. The first two modes shown here are primarily rotational oscillations. The third mode is the `bounce' mode of the IP-legs. All three oscillations are caused by a combination of vertical deformation of the IP-legs and bending of the spring-box. Inertial dampers can be placed in anti-nodes of the oscillations in order to effectively remove energy from these modes.}
	\label{springboxmodes}
\end{figure}

Due to the finite vertical compliance of the IP legs, three additional modes are observable in the system, with the spring-box bouncing vertically, and rotating in pitch and in roll. These modes have natural frequencies between 30 and 50\,Hz. 
 The finite element simulation (figure~\ref{springboxmodes}) shows the corresponding spring-box mode shapes.  It also shows that even though the thin upper flexures (operated in tension) and the thin walls of the IP-legs dominate the vertical compliance, the bending of the spring-box is a significant contribution. We investigated two approaches in order to improve the performance in the frequency band above 30\,Hz: the first was to change the geometry of the IP-legs to stiffen them in the vertical direction (see section~\ref{IPredesign}). This, in combination with spring-box stiffening will shift the vertical spring-box modes to higher frequencies and thereby widen the isolation frequency window. The second approach was to passively damp the resonances by installing dedicated inertial damping units (see section~\ref{springboxdamper}).

\FloatBarrier			

\subsection{IP-legs re-design}
\label{IPredesign}
The vertical IP-leg compliance can be improved by a redesign. However, there are several requirements for a new IP-leg design. Primarily, there must be a substantial increase in vertical stiffness without compromising the horizontal compliance. The new design must be compatible with the existing mechanics and attachment mechanisms, allow for the same range of horizontal motion (about one centimeter), and made from materials that can be machined and modified in-house. Additionally, the center of percussion compensation system (a bell-shaped addition to the base of the leg, see~\cite{wanner2013seismic}) must be able to compensate for the new mass and mass distribution of the leg.

The original IP-leg flexures were made from maraging steel, a material chosen because it combines very high strength (1.94\,GPa ultimate tensile strength), very low creep, and a low loss angle~\cite{beccaria1998creep, losurdo1999inverted}. However, maraging steel has a long lead-time for procurement. For the new design we substituted this material with Titanium grade 5. It is easy to machine, has a low internal loss and high strength~\cite{titanium, maraging}. Finite element simulations of the new IP-leg design show that displacing the payload by 10\,mm results in a maximum bending stress of approximately 60\,\% of the yield. 
Although this is much closer to yield than equivalent maraging steel flexures (which were at approximately 20\,\% of yield), the margin is still sufficient  considering that in normal operation the payload translates less than 1\,mm.

\begin{figure}[!htb]
	\centering
	\includegraphics[width=0.85\textwidth]{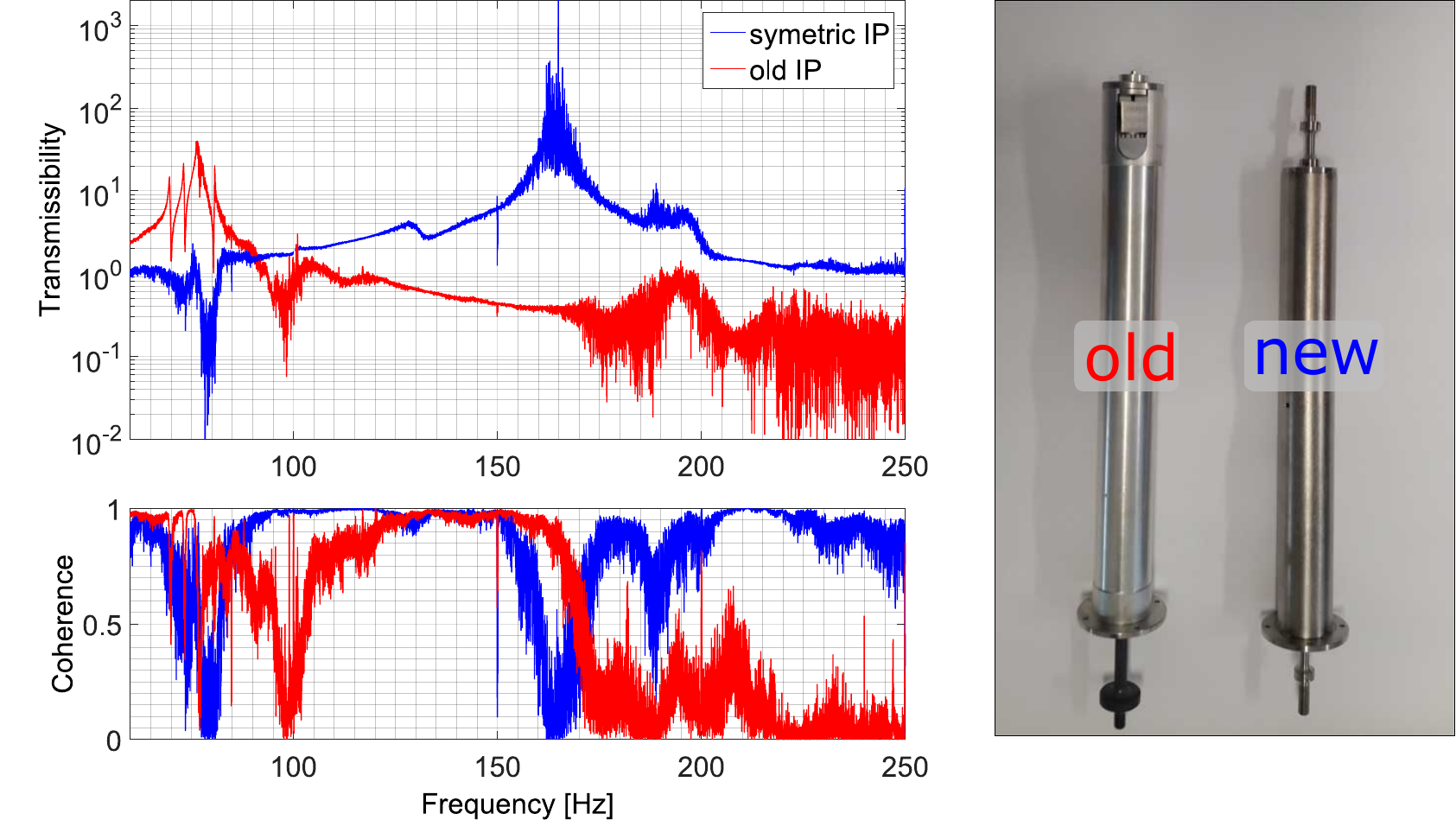}
	\caption{(left) Vertical transmissibility of old and new IP-legs: The improvement in vertical stiffness from the new design was experimentally tested by measuring the fundamental bounce mode of both IP-legs. Each leg was loaded with a $40\,$kg mass and the transmissibility was measured. The measurement result shows that the stiffness of the new design is more than four times higher than the original design. The dip in the symmetric IP-leg curve at $80\,$Hz and in the old IP-leg curve at $98\,$Hz is caused by tilt of the suspended mass. Due to geometric constraints, it was mounted on the two IP-Legs in different ways.\\ 
(right) Old and new IP-legs: In the original design the spring-box was suspended in tension from a thin upper flexure and the leg is a thin aluminum tube. The new design achieves greater vertical stiffness by using identical flexures at the top and the bottom and a thicker steel tube.} 
	\label{newip}
\end{figure}

The flexure thickness and shape were determined iteratively by finite element simulations and measurements to achieve a horizontal stiffness closer to the desired value. With $10\,$mm thick and $24.7\,$mm long flexures a horizontal stiffness of 11150\,N/m was achieved. This corresponds to a maximum load of 511\,kg. The leg itself is made from a stainless steel tube with $2\,$mm wall thickness, making it much stiffer than the original $1\,$mm wall thickness aluminum leg. The center of percussion compensator can be attached to the new design in the same way as the original. A photograph of the old and new IP-leg without the center of percussion compensator is shown in figure~\ref{newip}.
 
The vertical stiffness of the new IP-legs was measured in an auxiliary experiment, where they were were individually loaded with the same mass. The vertical transmissibility was measured by placing one geophone on the ground and one on the suspended mass. We determined their vertical stiffnesses from the fundamental bounce resonance of the IP-leg. Figure~\ref{newip} shows the vertical transmissibility of both IP-leg designs. The bounce mode of the old and new legs are 76\,Hz and 164\,Hz respectively. The resulting vertical spring constant of the old IP-leg is $9.1\times10^6\,$N/m, 8\,\% less than the results from the finite element simulation. The new symmetric IP-leg is more than 4 times stiffer, $42.4\times10^6\,$N/m. In combination with a planned stiffening of the spring-box, this will increase the frequency of the spring-box modes. If the spring-box was assumed to be a rigid body, the bounce mode on the new IP-legs would be about 100\,Hz, although the bending of the spring-box will lower the mode frequency. Preliminary simulations of the spring-box, that include additional stiffening structures, predict that the lowest vertical and vertical-tilt spring-box modes will be between 70\,Hz and 80\,Hz.
 
The new symmetric IP-legs in combination with spring-box stiffening will be implemented in the third AEI-SAS, which is currently under construction. The two other systems are already installed in the vacuum system and populated with electronics and optics. Replacing the old IP-legs is not feasible. However, the effect of the vertical spring-box modes can be mitigated by means of tuned dampers.

\FloatBarrier	
\subsection{Spring-box damping}
\label{springboxdamper}

Damping geometries, such as resonance dampers (\cite{wanner2013seismic,blomlvc}), can be retro-fitted into the AEI-SAS. A resonant damper is a secondary harmonic oscillator with large internal damping that is tuned to the resonance frequency of a primary oscillator. 

For practicality, the mass of the secondary oscillator is often much smaller than the mass of the primary oscillator. By placing it at an anti-node of the (primary) oscillation to be damped, the damper absorbs and dissipates energy through its internal loss mechanisms. If the mass ratio of the two oscillators is close to one, all resonances above the eigenfrequency of the damper can be efficiently damped. In this case the damper's inertia keeps it relatively stationary with respect to the primary oscillator, again dissipating energy via differential motion.
Anti-nodes of the spring-box oscillations were determined from finite element simulations (see figure~\ref{springboxmodes}). These are compared with experimental measurements of vertical vibration in the spring-box, made by placing a vertical geophone at potential damper positions.
\begin{figure}[!htb]
	\centering
	\includegraphics[width=0.85\textwidth]{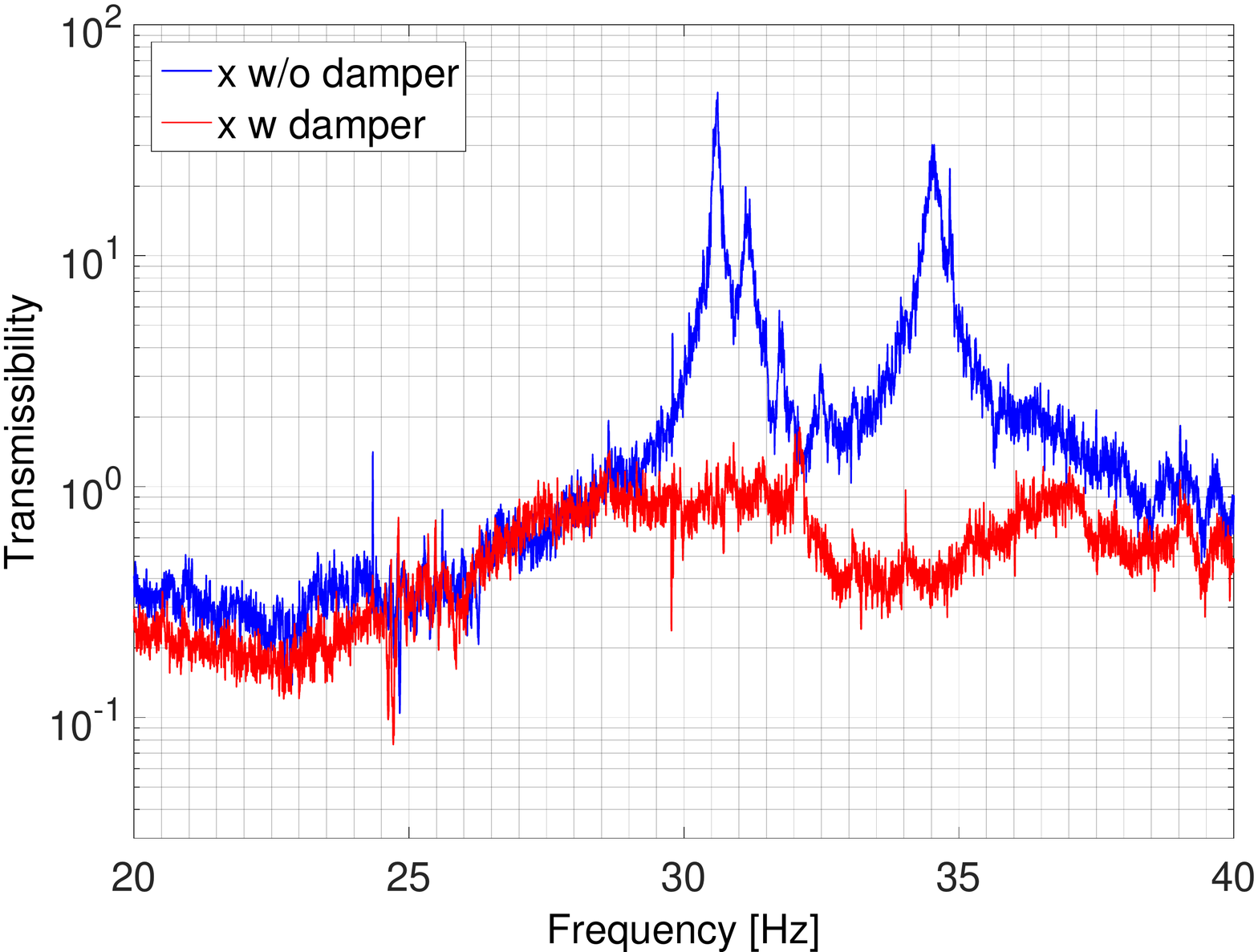}
	\caption{The transmissibility of horizontal ground motion to horizontal payload motion with (red) and without (blue) spring-box dampers. The payload motion was measured with an auxiliary accelerometer mounted directly on the payload. By placing three damping structures inside the spring-box, the three lowest spring-box modes were strongly damped.}
	\label{damper1}
\end{figure}

\begin{figure}[!htb]
	\centering
	\includegraphics[width=1\textwidth]{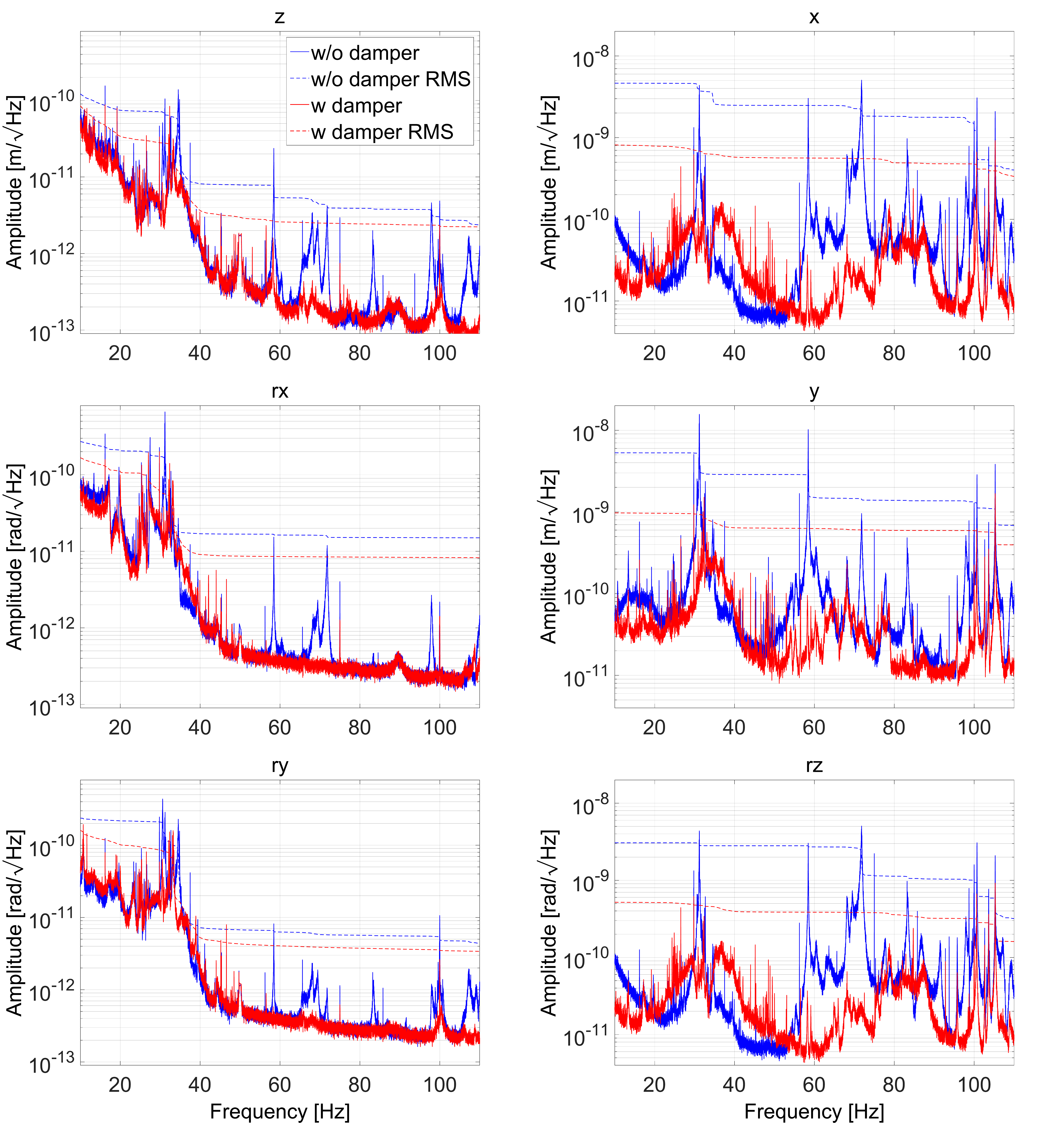}
	\caption{Spring-box damper performance in all 6 degrees of freedom. The plots on the left of this figure show the vertical payload motion (measured with the payload geophones) with (red) and without (blue) spring-box dampers. The plots on the right show the horizontal spring-box motion (measured with the three spring-box accelerometers) again with (red) and without (blue) spring-box dampers. Not only are the three fundamental spring-box modes between $30\,$Hz and $35\,$Hz damped, but higher frequency resonances are also significantly reduced. The RMS motion, integrated from high frequencies to low frequencies (dashed lines), is substantially reduced over a large frequency band. }
	\label{damper2}	
\end{figure}

Our dampers were $5\,$kg stainless steel masses placed on three Viton cylinders cut from O-ring seals. The lengths of the cylinders determines the resonance frequency, and the lossy Viton efficiently dissipates energy. A set of three such dampers was placed inside the spring-box. The improvement in table performance is shown in figure~\ref{damper1} and \ref{damper2}. As an example of the damping of spring-box modes, figure~\ref{damper1} shows the transmissibility from the ground to the spring-box in the horizontal direction. The transmissibility is reduced by a factor of up to  75 for the third mode. Resonances at higher frequencies are also strongly damped. Figure~\ref{damper2} shows the payload motion (z, rx, ry degrees of freedom) and the spring-box motion (x, y, rz degrees of freedom). The RMS motion is substantially reduced over a wide frequency band in all degrees of freedom.

\section{Conclusion}
We measured the performance of a seismic attenuation system that uses anti-springs to create very low-frequency resonances. The performance matches ideal models very well up to the first internal resonances. A peak isolation performance of $2.6 \times 10^{-3}$ was measured at 4\,Hz in the horizontal direction. In the vertical direction the ground motion is reduced by a factor of $2.4 \times 10^{-3}$ at 9\,Hz.
A horizontal fundamental resonance of the AEI-SAS of 0.1\,Hz enables a substantial reduction in motion in the frequency band near the mirror suspension resonances. At the lowest suspension resonance, 0.64\,Hz, the horizontal payload motion is 12 times lower than ground motion.
This enables a suspension design with low force actuators and passive eddy-current damping.

The internal resonances of the system were characterized by a combination of measurement, analytical modeling, and finite-element analysis. Following means of mitigating the effect of these resonances were proposed and tested:
\begin{itemize}
\item Fluorel pads decouple the spring-box and the payload. 
\item Spring-box dampers improve the isolation performance between 10 and 100\,Hz significantly.
\item New symmetric IP-legs which are 4 times stiffer than the old design will shift parasitic spring-box modes to higher frequencies, thus broaden the AEI-SAS's isolation window.     
\end{itemize}
The first two design changes were suitable to be retro-fitted into the first two AEI-SASs. All three presented improvements will be incorporated in the third system.
By shifting and damping the internal resonances, all mirror suspension resonances are now well isolated from ground motion.

\section{Acknowledgments}
The authors acknowledge support from the International Max Planck Research School (IMPRS) on Gravitational Wave Astronomy and from QUEST, the Center for Quantum Engineering and Space-Time Research.

\FloatBarrier		
\section*{References}	
\bibliography{literatur}

\end{document}